\documentclass[nonacm,sigplan]{acmart}
\usepackage{subfig}
\usepackage{graphicx}
\usepackage{subcaption}
\usepackage{balance}

\begin{document}

\title{The Interplay of Computing, Ethics, and Policy in Brain-Computer Interface Design}

\author{Muhammed Ugur}
%\email{muhammed.ugur@yale.edu}
\affiliation{
  \institution{Yale University}
  \postcode{06511}
  \country{}
}
\author{\centerline{\mbox{Raghavendra Pradyumna Pothukuchi}}}
%\email{}
\affiliation{
  \institution{Yale University}
  \postcode{06511}
  \country{}
}
\author{Abhishek Bhattacharjee}
%\email{abhishek.bhattacharjee@yale.edu}
\affiliation{
  \institution{Yale University}
  \postcode{06511}
  \country{}
}

\vspace{2mm}

\begin{abstract}

Brain-computer interfaces (BCIs) connect biological neurons in the brain with external systems like prosthetics and computers. They are increasingly incorporating processing capabilities to analyze and stimulate neural activity, and consequently, pose unique design challenges related to ethics, law, and policy. For the first time, this paper articulates how ethical, legal, and policy considerations can shape BCI architecture design, and how the decisions that architects make constrain or expand the ethical, legal, and policy frameworks that can be applied to them. %\abhishek{Let's say that while some recent work has begun looking at ethical and policy implications of BCIs -- cite the National Academies of Science work on this (https://www.nationalacademies.org/our-work/brain-machine-interface-technologies-scientific-technical-ethical-and-regulatory-issues-a-workshop) -- this is the first study that looks at what computer architects can do.} We envision this paper to stimulate broader conversations on this theme. \abhishek{Get rid of the last sentence.}

\end{abstract}

\maketitle % should come after the abstract
\pagestyle{plain} % should come right after \maketitle

\section{Introduction}

Brain-computer interfaces (BCIs) are devices that connect biological neurons in the brain with external systems like prosthetics and computers. These systems are advancing our understanding of the brain, helping treat many diseases and restore lost sensorimotor function~\cite{lebedev_brain-machine_2017}. 
They enable novel forms of human-machine interaction, and are being used in augmented reality/virtual reality (AR/VR) systems and industrial robotics~\cite{bciHci,muhl:survey,bciRobotics}.
%By enabling novel forms of human-machine interaction, they are also being found in augmented reality/virtual reality (AR/VR) systems, and industrial robotics~\cite{bciHci,muhl:survey,bciRobotics}.% \abhishek{How are they being found in AR/VR and industrial robotics? The last sentence is ambiguous.}

BCIs can sense or stimulate neural activity in the brain using several methods~\cite{lebedev_brain-machine_2017}. Some of these are invasive, i.e., they require surgery to place electrodes on the surface of the brain or inside. There are also non-invasive methods, e.g., those that use electrodes placed on the scalp, or which use methods like functional near-infrared sensing (fNIRS) to sense neural activity without surgery. 

Surgically implanted BCIs collect the highest fidelity signals with high spatio-temporal resolution~\cite{eegSignalCompare}, and hence, are mostly used in cutting-edge research to understand the brain, its diseases, and provide treatment. On the other hand, non-invasive methods pose lesser risk and are more broadly used, although the quality of signals they collect is low.
%\abhishek{The first sentence of a paragraph should be a topic sentence in conveying the point that you are trying to make. Yours seems to be that although there are invasive and non-invasive ways for BCIs to read/stimulate the brain, the invasive ways tend to give you the best signal fidelity and are therefore at the forefront of cutting-edge BCI applications. If that's the case, say that explicitly before going deeper into the different BCI technologies that one can use (which is what the rest of the paragraph is about).}

Today, we see an explosive growth in the use of BCIs, both implanted and wearable. There are several implanted BCIs undergoing clinical trials, with some already having received clinical approval~\cite{neuropace:rns}. There are also several wearable BCIs that are publicly available for purchase, yet regulatory agencies like the United States Food and Drug Administration (FDA) have not issued guidance on all such systems~\cite{tdcs}. 

As BCI usage grows, there is an increasing demand to integrate processing on board. Consequently, many researchers have responded by developing new architectures and circuits for on-device BCI processing~\cite{karageorgos:halo,Sriram2023,neuralDust}. 

BCI processor designers, however, confront unconventional design issues related to ethics, law and policy, in addition to difficult technical constraints. These novel issues arise due to the possibility of surgical implantation and the unique nature of these devices to directly sense neural and cognitive activity that can be highly revealing.
%, and additionally being surgically implanted for certain BCIs. 
Unfortunately, there is a lack of robust understanding of these issues and little guidance from policymaking agencies. Earlier discussions on the ethical and legal implications of BCIs have not considered the consequences that architectural choices, like on-device computing, can have on patients and clinical practice~\cite{ethicsWorkshop2023}.  %\abhishek{What discussions are you referring to?}

This paper presents how ethical and policy issues arise in making some of the most straightforward and basic choices in architecture design, which in turn affect device usage and the nature of policies and legal requirements that the system can support. Our goal is to inform and initiate conversations among a broader group of experts including computer architects, policymakers, legal scholars, and the various stakeholders including users, clinicians, and scientists. 

\section{Ethics, Policy, and Architecture Design}

Here we present the interplay between ethics, policy, and architecture design along a few crucial dimensions.

\noindent
\textbf{Specialization}: Historically, the choice of specialization vs. flexible processing has been made based on measures like power, performance, area, and cost. A specialized processor is more energy efficient, results in lower thermals, and prolongs lifetime when working with limited power supply, such as batteries. These traits are important for BCIs because they ultimately benefit the user. However, device specialization can limit support for newer versions of treatment methods, or treatments for newer conditions that the user may develop. As a result, the user might have to be undergo additional surgical processes for replacement and upgrades. While flexible processing avoids this issue, it comes with additional area and energy costs, which in turn, are not helpful for the user.  

There is a need for guidance from regulatory frameworks on what spectrum of architectures should be developed to offer patients the diversity of choice and meet their rights. For example, one possibility might be to build processors with a customized pool of accelerators for individuals to balance efficiency with flexibility. %These decisions would also shape the evolution of BCI processor market forces. 
Such a framework would also apply to other systems like mobile devices, but BCIs have a much tighter design space and higher impact on users. %the application  We see similar debates around obsoloscence, rbeginning to appear for conventional processors like those in mobile systems, but the tradeoffs have a drastic impacy for wearable or implanted nature of BCIs makes the tradeoffs

Furthermore, guidance is also necessary on the safety evaluation of generalizable implanted processors. Currently, the FDA approves devices for specific treatment applications. A flexible device, by its nature, can be used beyond its primary objective. Regulatory guidance helps determine what safeguards must be placed in the system. At the same time, technical design constraints like power and area also influence the nature of safeguards that can possibly be implemented.  

\vspace{1mm}
\noindent
\textbf{Upgradeability:} Closely related to the above challenge is the issue of upgradeability. Currently, we are unaware of minimum upgradeability or compatability requirements that BCI hardware or software must meet. Unlike a smartphone or even a cardiac pacemaker that is implanted, BCIs are active processing elements intricately tied to the neural and mental capabilities of an individual. As a result, when new versions of an implanted processor is released, we could have different classes of individuals with various capabilities owing to the different processor versions they would have. It is crucial to set minimum compatibility standards, and develop a formal policy framework to protect user rights. This too requires coordination between regulators and system designers.

\vspace{1mm}
\noindent
\textbf{Standards:} BCI standards can guarantee minimum functionality, and compatability for interoperability. There have been recent cases~\cite{Drew2020, BionicEye2022, Abandoned2022, HamzelouExplant2023} where BCIs implanted for individuals participating in clinical trials were forced to undergo explantation because the manufacturers went out of business. One issue prompting this drastic measure is that of device liability in the absence of the original manufacturer. However, even when the patients would assume all risk (as some individuals offered), there are device maintaineance and replacement concerns. This situation is not unique to implanted BCIs, since there are many non-invasive BCIs too, which are being explored for treatment. BCI standards could help address some of these challenges. 

BCI standards could be defined for the various hardware components, such as the power delivery systems, processor, sensors and communication modules. %Such standards would also help when it is necessary for multiple heterogeneous implants to be used to support an application. 
Such standards could also apply to software frameworks that manage the BCI. Supporting standardized interfaces, however, inevitably requires additional processing, resulting in additional power dissipation that might not be desirable for a user. Some of the standards might not even be feasible to implement in the limited energy or power budget of the devices. Thus, regulatory recommendations are required to guide architectures that balance patient interests, rights and architectural feasibility.

\vspace{1mm}
\noindent
\textbf{Security, access, and autonomy:} The sensitivity classification of neural data, its access, and methods of protection are all currently being explored~\cite{WajnermanPaz2021,Ryberg2016}. However, it is important to consider the role of architecture in these decisions. 

Consider encryption, for example. Supporting encryption requires energy and dissipates heat. This impacts brain physiology, and also limits the other applications that can be run simultaneously. It is important to specify recommendations on when and how to balance encryption with the BCI's primary applications. Ad hoc measures are undesirable since these decisions have a direct impact on the user's life and even when safe, may violate a user's preferences or rights.% certain device functions, and if it has to occur, the contexts in which it should take place (e.g., when a BCI is running low on power). 

An important related aspect to consider is the ownership of the device, neural data, and mechanisms of data sharing. Provisions must be made in the hardware or software to support any of these features. For example, without explicit support in the hardware for a user to authenticate with the device, they might not be able to access their data even if they were given the right to do so at a later stage. However, all these mechanisms also impact power, performance and area. Thus, architects must be involved closely with regulators, ethicists, and policymakers to determine neural data rights, ownership, access protocols, and user autonomy. 

Lastly, the autonomy of the BCI device itself is under exploration. It is possible for a BCI to learn of the intent to perform malicious thoughts and acts~\cite{Scheibner2021}. It is not clear whether the BCI should log or report such events, or ignore them entirely. Whether a BCI can act on its own is not limited to these situations. Consider a patient experiencing a debilitating condition that the BCI becomes aware of, but which was not the primary target of the BCI. It might be possible to save the user's life by reporting this event to a doctor. All of these possibilites could be addressed, but require system support and close coordination between regulatory bodies and architects to arrive at appropriate frameworks. 

\textit{Remarks}: The interplay of ethics, policy, and computer architecture in BCI design that we presented here is by no means complete. Our goal has been to emphasize the need for BCI architects to become aware of these novel challenges, and engage with appropriate experts \textit{at design time}. Several issues (e.g., data protection, upgradeability) are applicable for other electronic devices too, and have been the target of legislative measures. However, these issues are uniquely severe for BCIs.% the problem for BCI architects.  

\section{Related Work}

A recent meeting sponsored by the United States national academies raises several issues on the ethical, regulatory and policy implications of BCIs~\cite{ethicsWorkshop2023}. However, the discussion did not consider BCIs that include on-device processing, which allows more complex processing and autonomous usage. 

Many studies argue for neural data protection and individual privacy using encryption methods and HIPAA compliance~\cite{Ienca2022,Agarwal2019,Thapa2021,YustePrivacy2023}. However, as we have presented above, encryption is not a panacea, and supporting itself requires balancing several competing interests.

\section{Conclusion}

This paper described some ways in which computer architecture design interacts with ethical, law, and policy considerations in the BCI domain. By no means is our work complete. However, it is urgent to initiate conversations around these themes and develop regulatory guidance.

\balance
\bibliographystyle{ACM-Reference-Format}
\bibliography{references}

\end{document}